# Low-Temperature Heat Capacity and Phonon Dynamics in Expanded Graphite and EG–MWCNTs Composites

A.I. Krivchikov[1,2], A. Jeżowski[2], M.S. Barabashko[1], G. Dovbeshko[2,3], D.E. Hurova[1], N.N. Galtsov[1], V. Boiko[2,3], Yu. Sementsov[4], A. Glamazda[1], V. Sagan[1], Yu. Horbatenko[1], O.A. Korolyuk[1], O.O. Romantsova[1,2], D. Szewczyk[2]*

[1]B. Verkin Institute for Low Temperature Physics and Engineering, NAS of Ukraine, Nauky Ave. 47, 61103 Kharkiv, Ukraine

[2]Institute of Low Temperature and Structure Research PAS, Okólna 2, 50-422 Wrocław, Poland

[3]Institute of Physics, NAS of Ukraine, Prospect Nauky, 46, Kyiv 03028, Ukraine

[4]Chuiko Institute of Surface Chemistry, NAS of Ukraine, General Naumov str. 17, Kyiv, 03164, Ukraine

*corresponding author. d.szewczyk@intibs.pl

## Abstract

The specific heat of expanded graphite (EG) and EG–multiwalled carbon nanotube (MWCNT) composites (1.0 and 3.0 wt.% MWCNTs) was measured between 2 and 300 K. The low-temperature heat capacity is dominated by out-of-plane flexural phonons with quadratic dispersion, characteristic of two-dimensional layered systems. Compared with crystalline graphite, EG exhibits enhanced heat capacity due to increased defect density and reduced interlayer coupling. Structural characterization (XRD, Raman, EDS) confirmed variations in stacking order and defect concentration. The data were fitted using a three-term model ($C_1T + C_3T^3 + C_5T^5$), where the negative $C_5$ term indicates quadratic phonon dispersion. The results demonstrate the influence of MWCNT integration and structural disorder on phonon dynamics and anisotropic heat capacity in EG-based composites.

### 1. Introduction

The low-temperature heat capacity ($C_p$) of nonmetallic crystalline solids with ideal three-dimensional structures typically follows a cubic temperature dependence ($\sim T^3$), a direct consequence of the quadratic increase in the density of acoustic phonon states with frequency, as described by the Debye lattice dynamics model [1]. In contrast, disordered systems such as structural glasses, despite exhibiting isotropic elastic behavior, display significantly higher specific heat compared to their crystalline counterparts. This excess specific heat, evident in deviations of the reduced heat capacity ($C_p/T^3$) versus temperature, reflects additional low-energy vibrational excitations beyond the Debye framework [2].

At temperatures below 1 K, disordered dielectric solids frequently display an extra quasi-linear term in $C_p$. This term is generally attributed to a constant density of tunneling states or two-level systems (TLS), as described by the Standard Tunneling Model (STM) [3, 4]. For carbon-based materials, phonon excitations predominantly govern the low-temperature heat capacity, although theoretical descriptions vary widely across different carbon allotropes. Pristine graphite adheres to the Debye model, exhibiting $T^3$ dependence characteristic of a highly ordered phonon spectrum. In contrast, reduced-dimensional and structurally disordered carbon forms—such as graphene nanoplatelets (GNPs), reduced graphene oxide (rGO), and carbon nanotubes (CNTs)—deviate significantly from this behavior. For example, isolated CNTs, dominated by one-dimensional phonon transport, may exhibit a linear ($T$) dependence, while few-layer graphene systems tend to follow an intermediate $T^2$ scaling, indicative of two-dimensional vibrational dynamics [5]. Expanded graphite and mesophase pitch-based carbon fibers (MPCFs) often display strong anisotropy, defects, and grain boundary effects that introduce additional low-frequency vibrational modes, thereby modifying the temperature exponent and magnitude of specific heat [6].

For expanded graphite composites incorporating MWCNTs, both density and disorder play crucial roles in determining low-temperature heat capacity. Structural disorder—including defects, grain boundaries, and morphological inhomogeneities—introduces additional vibrational states, leading to anomalies in $C_p$ [7-18]. The morphology of EG, characterized by defects, porosity, and partial reduction, further amplifies phonon scattering. Such deviations from the Debye law offer crucial insights into phonon dynamics and the role of disorder in determining thermal behavior. A thorough understanding and accurate modeling of these effects are essential for optimizing the thermal performance of these composites in advanced thermal management applications.

Expanded graphite composites with controlled density and functionalization present promising opportunities for hydrogen and thermal energy storage, fuel cells, batteries, supercapacitors, and biosensors [19]. In carbon nanotechnology, the enhancement of thermal conductivity, stability, and heat transfer is of importance [20-24]. For example, the thermal conductivity of polymer matrices strongly depends on the morphology of embedded carbon fillers, a crucial factor in designing materials for heat exchangers and electronics.

Graphite's layered structure results in pronounced anisotropy in thermal and electrical properties [8, 25]. Its high in-plane thermal conductivity coupled with out-of-plane insulating behavior makes graphite-containing composites ideal for heat dissipation in electronics [26]. Additionally, graphite's chemical stability, low thermal expansion coefficient, and excellent thermal resistance make it highly valuable for high-performance thermal interface materials [27, 28].

Despite their potential, the low-temperature thermal and electrical properties of EG remain insufficiently explored. Similar challenges arise in other carbon nanomaterials—such as thermally reduced graphene oxide (trGO), graphene oxide-functionalized polymers, and multiwalled carbon nanotubes—where self-agglomeration often hinders functionality [22, 29-32]. For instance, CNT–CNT coupling can lower thermal conductivity by an order of magnitude [31, 33]. It is reasonable to assume that interlayer interactions in EG similarly influence phonon transport [32]. Recent studies have shown that integrating a controlled amount of MWCNTs into graphene-based materials can yield composites with enhanced surface area, improved chemical stability, and superior thermal and mechanical properties [34-36]. The overall performance of EG-MWCNTs composites in thermal interface applications strongly depends on both intrinsic material properties and thermal quality of their interfaces—parameters heavily influenced by fabrication methods [34].

Thus, further research into the low-temperature thermal behavior of EG-based composites is critical to fully elucidate phonon transport mechanisms and optimize material design for advanced thermal management applications. The novelty of this work lies in investigating the temperature-dependent thermal properties of unique low-dimensional carbon–carbon composites with varying densities, emphasizing their structural and interfacial interactions. The study integrates experimental analysis of EG and EG-MWCNTs composite heat capacity with complementary structural characterization via Raman spectroscopy and X-ray diffraction. A particularly intriguing aspect is the potential discovery of novel low-temperature heat capacity effects arising from the complex microstructure of these composites.

Experimental studies remain essential for refining theoretical models by providing direct data of heat capacity across carbon allotropes. Calorimetric experiments—combined with structural characterization techniques such as Raman spectroscopy, XRD, and electron microscopy—allow precise quantification of defect density, crystallite size, interlayer interactions, and morphological effects on phonon transport. These empirical insights not only enhance theoretical predictions, often based on idealized Debye or low-dimensional models, but also reveal additional contributions from disorder that are fundamental to understanding and engineering thermal properties in carbon-based nanomaterials.

The aim and novelty of this work is to investigate the physical phenomena underlying the temperature-dependent thermal properties of unique low-dimensional carbon–carbon composites with different defect densities, with particular attention to their structure and interphase interactions. In our study, we focus on an experimental analysis of the low-temperature heat capacity behavior of Expanded graphite (EG) and EG–multiwalled carbon nanotube (MWCNT) composites, in conjunction with structural characterization via Raman spectroscopy and X-ray diffraction (XRD). A particularly intriguing aspect of this research is the potential discovery of novel low-temperature heat capacity effects emerging from the composites' complex structure.

## 2. Materials and methods

*Fabrication of Carbon–Carbon Composites*

Expanded graphite (EG) composites incorporating multi-walled carbon nanotubes (MWCNTs) were synthesized using advanced exfoliation and intercalation techniques [35, 36]. EG is obtained by thermal shock treatment of hydrolyzed and dried graphite intercalation compounds obtained by oxidative intercalation process by chemical or electrochemical methods. This procedure results in significant delamination and highly porous nanostructure of expanded graphite. Characterized by slit-shaped and cylindrical defects (2–40 nm in cross-section), EG exhibits enhanced surface area and structural heterogeneity, confirmed through Raman spectroscopy, positron annihilation, and low-temperature nitrogen adsorption-desorption experiments [35, 37].

MWCNTs were synthesized via catalytic chemical vapor deposition (CCVD) [38]. These nanotubes exhibit an average diameter of 10–20 nm, a specific surface area of 200–400 $m^2/g$ (determined by argon desorption), and a bulk density of 20–40 $g/dm^3$.

*Preparation of EG-MWCNTs Composites*

EG-MWCNT composites were synthesized by chemical oxidation with a solution of potassium dichromate in 94% sulfuric acid (1.0 and 3.0 wt. %) or by anodic (electrochemical) oxidation in 94% sulfuric acid (1.0 wt. %) at an electrical quantity of 90–120 A·h/kg, combining carbon nanotube deagglomeration in concentrated acid and graphite intercalation [35, 36]. In both cases, first-order intercalation compounds (alternating layers of intercalant and graphite one after the other) were obtained. [35, 36]. After extensive washing (6–7 pH) and drying, the material was subjected to rapid thermal shock (850–1250°C), followed by rolling into foils with densities of 0.2–1.7 $g/cm^3$.

Two oxidative intercalation methods were used to obtain expanded graphite with varying degrees of structural disorder:

- **Chemical oxidation**: In addition to the graphite being intercalated using a solution of potassium dichromate in sulfuric acid, as mentioned above, which made it possible to obtain the intercalated compound of stage I, intercalation was carried out using ammonium persulfate ($(NH_4)_2S_2O_8$) as an oxidant. In this case, intercalation compounds of mixed stages II and III were obtained. Further hydrolysis (washing with distilled water on a filter to neutral pH), drying and heat treatment in thermal shock mode were carried out [35].
- **Electrochemical oxidation**: This method was used in two modes: 1 - using concentrated sulfuric acid (94%, iron electrodes), the amount of electricity passed is 90–120 A·h /kg. We obtain the I stage of the intercalated compound. 2 - using 55% $H_2SO_4$ (lead electrodes), the amount of electricity passed is 60–100 Ah/kg. We obtain a mixture of II and III stages of intercalation. Further hydrolysis, drying and heat treatment in the thermal shock mode give different structural states of expanded graphite [35].

The samples were classified based on different synthesis conditions and carbon nanotube (CNTs) incorporation. **Chem** and **EleChem** denote chemical and electrochemical oxidation, respectively, and we add the indices I and II-III, i.e., $\mathbf{Chem^{I}}$ and $\mathbf{Chem^{II\text{-}III}}$, $\mathbf{EleChem^{I}}$ and $\mathbf{EleChem^{II\text{-}III}}$, which means that the samples were oxidized chemically or electrochemically to the first stage of intercalation or a mixture of the second and third stages and correspond to different structural states of expanded graphite. Samples labeled as CNTs contained nanotubes (1.0, 3.0 wt.%). Samples were prepared with different oxidants (e.g., ammonium persulfate, potassium dichromate, which gave different stages of intercalated compounds) and CNTs concentrations that affected the final morphology and thermal properties. Commercial EG materials (SINOGRAF SA: GFL05, GFC025, GFC10) were included for comparative analysis.

*Heat Capacity Measurements*

Heat capacity was measured via thermal relaxation method using a Physical Property Measurement System (PPMS®) from Quantum Design Inc., operating in the Heat Capacity Option across a 2–300 K temperature range. Measurement errors were below 5% above 100 K and 2% below 50 K. Apiezon N thermal grease ensured thermal coupling between the sample and the measurement platform. Data analysis followed protocols established in [39, 40].

*Structural Characterization*

EG and EG-MWCNT composites were characterized via:

- **Energy-dispersive X-ray spectroscopy (EDS)**: Elemental composition were determined with a Field Emission Scanning Electron Microscope (FE-SEM, FEI Nova NanoSEM 230).
- **X-ray diffraction (XRD)**: Powder samples measured using a PANalytical X'pert Pro diffractometer (Cu-Kα radiation, λ = 1.5418Å, Bragg-Brentano geometry, room temperature).
- **Raman Spectroscopy**: Structural homogeneity assessment via a Renishaw in Via Raman microscope with a DM 2500 Leica confocal system (λ = 514 nm, CCD detection). Spectra were collected from multiple randomly selected spots, accumulated three times (10 s exposure), over a 1000–3200 $cm^{-1}$ range.

## 3. Results

The elemental composition of EG and EG-MWCNTs composites, determined via energy-dispersive X-ray spectroscopy (EDS), is summarized in Table 1. The analysis indicates a noticeable increase in oxygen, sulfur, and chromium content in EG samples synthesized through chemical oxidation compared to those obtained via electrochemical oxidation.

**Table 1.** EDS analysis for EG and EG-MWCNTs materials.

| № | Sample | Intercalant, $H_2SO_4$, concentration, % | oxidizer | C at% | O at% | S at% | Cr at% | K at% | other (Al, Si, Fe) at% |
|---|---|---|---|---|---|---|---|---|---|
| **EG** | | | | | | | | | |
| 1 | #1-Sinograf **GFL**05 ρ = 0.7 g/cm$^3$ | — | — | 94.97 | 4.8 | 0.18 | - | - | 0.05 |
| 2 | #2-Sinograf **GFC**025 ρ = 1.0 g/cm$^3$ | — | — | 94.61 | 5.2 | 0.15 | - | - | 0.04 |
| 3 | #3-Sinograf **GFC**10 ρ = 1.0 g/cm$^3$ | — | — | 94.38 | 5.4 | 0.16 | - | - | 0.06- |
| 4 | #4-**Chem$^{I}$** ρ = 0.235 g/cm$^3$ | 94 | $K_2Cr_2O_7$ | 92.84 | 5.52 | 1.25 | 0.33 | - | 0.06 |
| 5 | #5-**Chem$^{II-III}$** ρ = 0.231 g/cm$^3$ | 94 | $(NH_4)_2S_2O_8$ | 97.77 | 1.75 | 0.44 | - | - | 0.04 |
| 6 | #6-**EleChem$^{II-III}$** ρ = 0.357 g/cm$^3$ | 55 | current | 91.52 | 8.2 | 0.2 | - | - | 0.08 |
| 7 | #7-**Chem$^{I}$** ρ = 0.73 g/cm$^3$ | 94 | $K_2Cr_2O_7$ | 94.75 | 4.8 | 0.18 | 0.18 | 0.05 | 0.04 |
| 8 | #8-**Chem$^{I}$** ρ = 1.3 g/cm$^3$ | 94 | $K_2Cr_2O_7$ | 93.51 | 5.8 | 0.19 | 0.39 | 0.07 | 0.04 |
| **EG-1%MWCNTs** | | | | | | | | | |
| 9 | #9-**EleChem$^{I}$**-1%CNTs ρ = 0.357 g/cm$^3$ | 94 | current | 97.83 | 1.62 | 0.13 | 0.38 | - | 0.22 |
| 10 | #10-**Chem$^{I}$**-1%CNTs ρ = 0.245 g/cm$^3$ | 94 | $K_2Cr_2O_7$ | 95.49 | 3.56 | 0.27 | 0.37 | 0.11 | 0,1 |
| 11 | #11-**EleChem$^{I}$**-1%CNTs ρ = 0.233 g/cm$^3$ | 94 | current | 95.77 | 3.77 | 0.36 | - | - | - |
| **EG-3%MWCNTs** | | | | | | | | | |
| 12 | #12-**Chem$^{I}$**-1%CNTs ρ = 1.74 g/cm$^3$ | 94 | $K_2Cr_2O_7$ | 98.26 | 1.08 | 0.26 | 0.26 | 0.07 | 0.07 |
| 13 | #13-**Chem$^{I}$**-3%CNTs ρ = 0.966 g/cm$^3$ | 94 | $K_2Cr_2O_7$ | 97.86 | 1.68 | 0.15 | 0.15 | 0.07 | 0.09 |
| 14 | #14-**Chem$^{I}$**-3%CNTs ρ = 1.62 g/cm$^3$ | 94 | $K_2Cr_2O_7$ | 97.64 | 1.92 | 0.15 | 0.12 | 0.06 | 0.11 |

The Raman spectra of EG and EG-MWCNTs composites, presented in Fig. 1, exhibit characteristic G and 2D bands (the characteristic parameters of the typical bands are summarized in Table 2), which correspond to the in-plane stretching vibrations of $sp^2$ hybridized carbon atoms and the second-order response of the defect-induced D-band. Notably, the 2D mode is consistently observed even in the absence of the D mode, indicating that defect presence is not a prerequisite for two-phonon activation. To analyze the vibrational properties of the samples, the Raman spectra were fitted using Voigt functions, which represent a convolution of Lorentzian and Gaussian components. Additionally, the spectra reveal distinct D, D', and 2D bands associated with disorder-activated vibrational modes in microcrystalline graphite. Structural disorder in expanded graphite primarily originates from lattice defects, including crystallite boundaries, impurities, and edge effects, which significantly influence its vibrational characteristics.

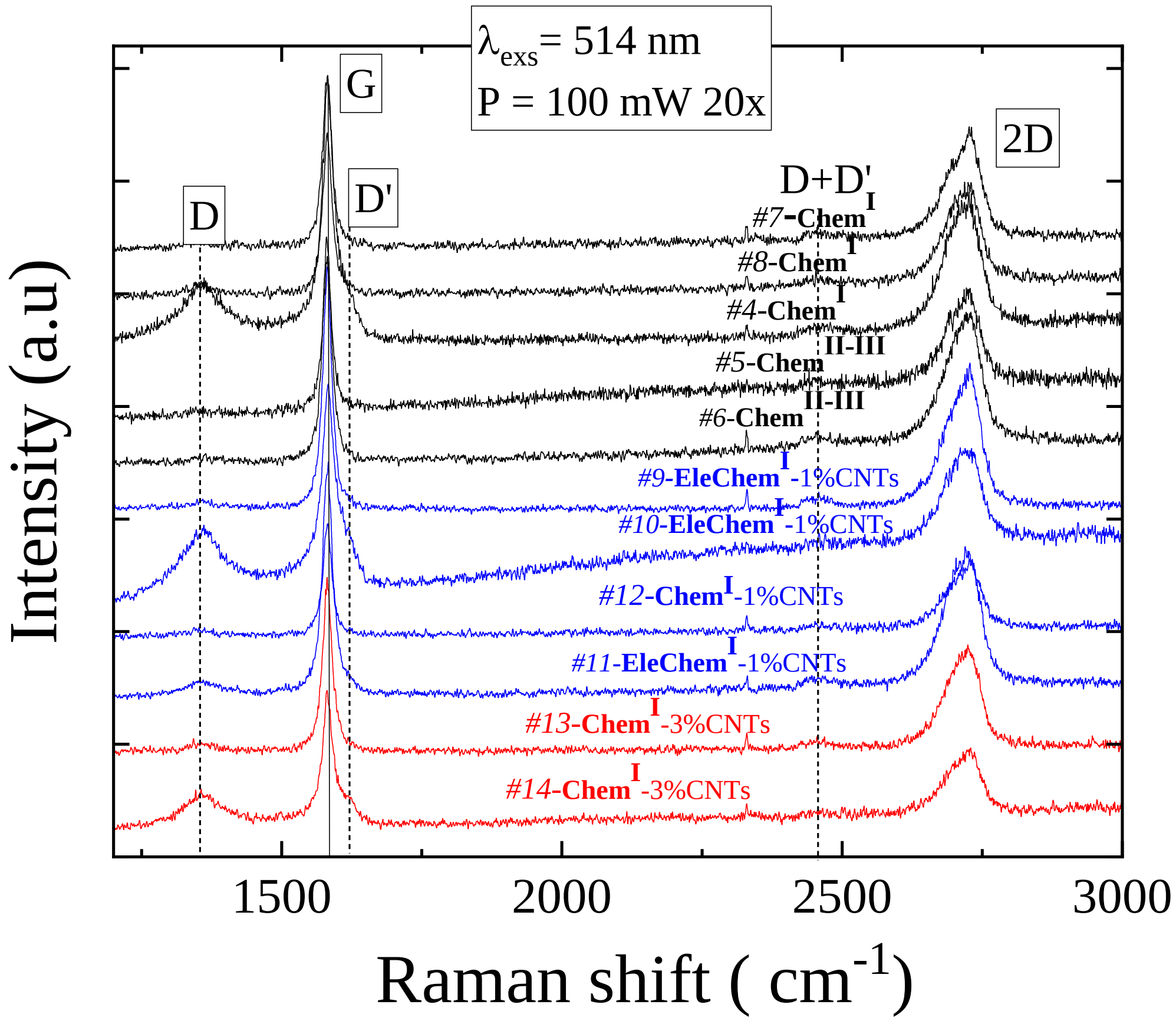

**Fig. 1** Raman spectra of EG and EG-MWCNTs composites.

Notably, the low-intensity D-band peak suggests a reduced defect density in the honeycomb lattice, as the D-band is typically associated with structural imperfections in graphene. These defects arise from disruptions in the hexagonal arrangement of $sp^2$ carbon atoms. The D' mode, observed around ~1620 $cm^{-1}$, originates from intra-valley defect scattering and serves as an indicator of point defects or edge irregularities. Additionally, the combination mode (2D') at ~2940 cm$^{-1}$ further confirms the presence of structural disorder.

The 2D band undergoes significant modifications depending on the thickness of AB-stacked flakes. In multilayer graphene and graphite, the 2D peak comprises a convolution of two sub-peaks, $2D_1$ and $2D_2$. The $I_{2D}/I_G$ ratio for expanded graphite is approximately ~ 1÷2 (Table 2), and its relatively low value, coupled with the characteristic double-hump line shape, is indicative of graphitic multilayers [41]. The $I_D/I_G$ ratio is widely used to assess in-plane structural order, as it inversely correlates with the average defect separation and the effective size of in-plane crystallites or graphitic domains [42-44].

The peak at ~1350 cm$^{-1}$, known as the disorder band (D-band), emerges due to interactions between sp$^2$ carbon rings and graphene edges or other network defects. The parameters of the Raman bands for EG and EG-MWCNTs composites align well with previous findings [36, 44, 45].

**Table 2.** Raman spectroscopy experimental data for EG and EG-MWCNTs composites, including the $I_{2D}/I_G$ ratio, which reflects the fraction of turbostratic single-layer graphene.

| **Line frequency in cm$^{-1}$** | | | | | | | | | | |
|---|---|---|---|---|---|---|---|---|---|---|
| Sample | D | A | G | D' | D+D' | $2D_1$ | $2D_2$ | 2D' | $I_{2D}/I_G$ | $I_D/I_G$ |
| **EG** | | | | | | | | | | |
| #4-**Chem$^{I}$** | 1358.9 | 1512 | 1582.6 | 1619 | 2463.8 | 2697.2 | 2730.4 | 3290.8 | 1.07 | 0.89 |
| #5-**Chem$^{II-III}$** | 1353 | | 1581.2 | | 2450.3 | 2699.3 | 2730.5 | 2865.6 | 1.67 | 0.07 |
| #6-**EleChem$^{II-III}$** | 1359.9 | | 1580.9 | | 2454.8 | 2697.8 | 2730.8 | 3218.2 | 2.17 | 0.06 |
| **#7-Chem$^{I}$** | 1358 | | 1586.5 | | | 2702.5 | 2730.5 | | 1.08 | 0.33 |
| **#8- Chem$^{I}$** | 1358.3 | | 1586.9 | | | 2699.8 | 2735.3 | | 1.08 | 0.33 |
| **EG – 1% MWCNTs** | | | | | | | | | | |
| #9-**EleChem$^{I}$** -1%CNTs | 1356.6 | | 1583 | 1618.5 | | 2698.6 | 2731.4 | | 1.25 | 0.06 |
| #10-**Chem**-CNT | 1352.6 | 1528.1 | 1581.3 | 1620.6 | 2457.8 | 2685 | 2722.4 | | 1.07 | 1.01 |
| #11-**EleChem$^{I}$** -1%CNTs | 1356.5 | 1518.4 | 1581.6 | 1616.1 | 2461.9 | 2698.6 | 2728.9 | | 1.95 | 0.26 |
| #12- **Chem$^{I}$** - 1%CNTs | 1352.1 | | 1581.3 | | | 2700.9 | 2732.3 | | 1.93 | 0.15 |
| **EG – 3% MWCNTs** | | | | | | | | | | |
| #13- **Chem$^{I}$** - 3%CNTs | 1358 | | 1581.5 | 1621.6 | 2456.2 | 2702.5 | 2730.5 | | 2.05 | 0.1 |
| #14- **Chem$^{I}$** - 3%CNTs | 1359.4 | 1536.6 | 1581.6 | 1616.6 | | 2697 | 2731.7 | | 1.51 | 1.31 |

Fig. 2 presents the X-ray diffraction (XRD) patterns of EG and EG-MWCNTs composites. The diffraction spectra exhibit reflections associated with the basal planes (00l) of the polycrystalline layered graphite structure. The coherent scattering region ($L_c$) was estimated using the Scherrer equation [46].

$$L_c = \frac{0.9 \cdot \lambda}{\beta \cdot \cos\theta},$$

where λ is the X-ray wavelength, $\theta$ is the Bragg angle of the 002 diffraction peak, and $\beta$ is the full-width at half-maximum (FWHM) of this peak.

Additional weak reflections accompanying the predominant hexagonal phase were detected in all low-density samples (ρ < 0.4 g cm$^{-3}$) within the small-angle region.

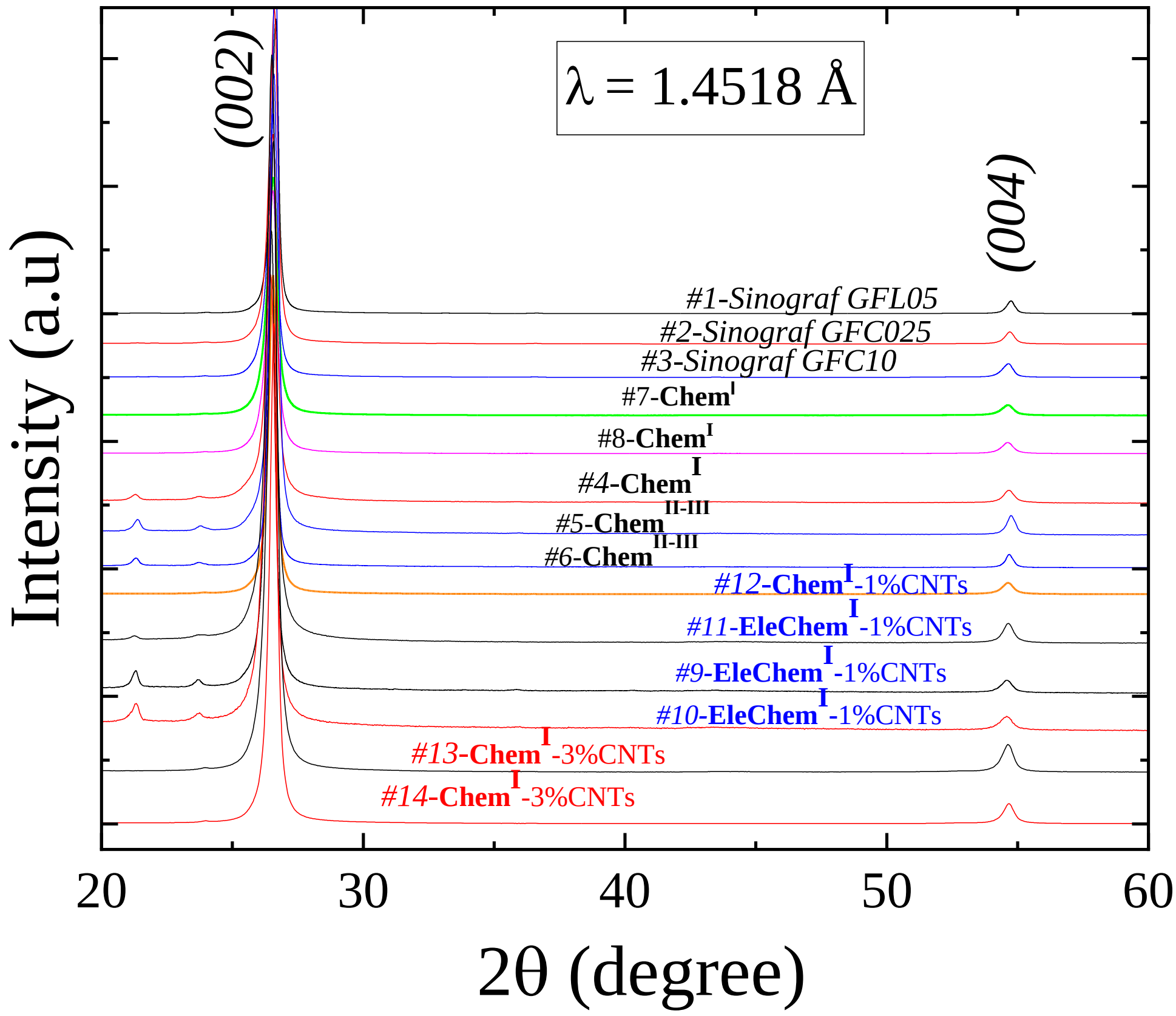

**Fig. 2** XRD patterns of EG and EG-MWCNTs composites including commercial Sinograf samples.

The position of the (002) reflection and the full width at half maximum (FWHM) of the (002) peak in the XRD patterns of EG and EG-MWCNTs composites exhibit variations of no more than 5% and 23%, respectively, from their average values (Table 3). The observed increase in FWHM upon CNT incorporation into the EG matrix can be attributed to multiple factors, including internal stresses within graphene layers, the intrinsically heterogeneous and porous structure of EG, and the initial $sp^2$ defectiveness that arises from the synthesis process of EG-MWCNTs composites.

**Table 3.** X-ray diffraction analysis data for EG and EG-MWCNTs composites, including structural parameters and crystallographic properties. CNT concentration in the composites is provided by the material manufacturer. The interplane distance *d* and the size of coherent scattering blocks $L_c$ were are calculated on the basis of the Bragg's law $d = \lambda/2\sin\Theta_{002}$ and Sherrer equation.

| Sample | *hkl* | 2θ, degrees | FWHM, degrees | d, Å | $L_c$, nm |
|---|---|---|---|---|---|
| **EG** | | | | | |
| #1-Sinograf **GFL**05 | 002 | 26.60 | 0.31 | 3.35(4) | 26 |
| | 004 | 54.69 | 0.41 | | |
| #2-Sinograf **GFC**025 | 002 | 26.58 | 0.36 | 3.35(5) | 22 |
| | 004 | 54.67 | 0.40 | | |
| #3-Sinograf **GFC**10 | 002 | 26.57 | 0.35 | 3.35(8) | 23 |
| | 004 | 54.62 | 0.46 | | |
| **#4-Chem$^{I}$** | 002 | 26.52 | 0.46 | 3.35(9) | 18 |
| | 004 | 54.66 | 0.42 | | |
| #5-**Chem$^{II-III}$** | 002 | 26.60 | 0.38 | 3.35(4) | 22 |
| | 004 | 54.75 | 0.44 | | |
| #6-**EleChem$^{II-III}$** | 002 | 26.54 | 0.32 | 3.35(8) | 26 |
| | 004 | 54.68 | 0.39 | | |
| #7-**Chem$^{I}$** | 002 | 26.52 | 0.43 | 3.36(1) | 19 |
| | 004 | 54.61 | 0.49 | | |
| #8- **Chem$^{I}$** | 002 | 26.51 | 0.42 | 3.36(2) | 19 |
| | 004 | 54.61 | 0.46 | | |
| **EG – 1% MWCNTs** | | | | | |
| #9-**EleChem$^{I}$**-1%CNTs | 002 | 26.46 | 0.46 | 3.36(6) | 21 |
| | 004 | 54.56 | 0.45 | | |
| #10-**Chem$^{I}$**-1%CNTs | 002 | 26.44 | 0.41 | 3.36(9) | 20 |
| | 004 | 54.56 | 0.54 | | |
| #12- **Chem$^{I}$** -1%CNTs | 002 | 26.53 | 0.33 | 3.36(0) | 25 |
| | 004 | 54.62 | 0.38 | | |
| #11-**EleChem$^{I}$**-1%CNTs | 002 | 26.48 | 0.38 | 3.36(4) | 21 |
| | 004 | 54.62 | 0.43 | | |
| **EG – 3% MWCNTs** | | | | | |
| #13- **Chem$^{I}$** -3%CNTs | 002 | 26.54 | 0.37 | 3.36(0) | 22 |
| | 004 | 54.61 | 0.48 | | |
| #14- **Chem$^{I}$** -3%CNTs | 002 | 26.56 | 0.41 | 3.35(8) | 24 |
| | 004 | 54.63 | 0.8 | | |

For the six less dense samples ($\rho$ < 0.36 g/cm³), an additional peak at 2θ ≈ 21.3° is observed, indicating the presence of a substructure within these highly heterogeneous porous composites, which are possibly due to the ordered distribution of intercalant residues (p.102 in [35]). Raman spectroscopy further reveals distinct differences in the defect and turbostratic structures of EG and EG-MWCNTs composites synthesized via chemical and electrochemical oxidation, demonstrating the influence of synthesis conditions on the degree of disorder and overall material properties.

Additionally, elastic–plastic deformation of the samples was carried out using a hydraulic press. Stacks of 5 × 5 mm² sheets were subjected to a pressure of 2 MPa. It induces both reversible (elastic) and irreversible (plastic) modifications within the worm-like layered structure of low-density samples ($\rho$ ~ from 0.2 to 0.4 g/cm³), thoroughly investigated in [47]. In addition, [48] reported that pressing expanded graphite powder is accompanied by a decrease in paramagnetic centre content by several orders of magnitude. This also qualitatively confirms significant restructuring of the material's structure.

Figures 3 and 4 illustrate the temperature dependence of heat capacity, $C_p(T)$, for expanded graphite (EG) samples, compared to crystalline graphite, across the 2–300 K temperature range [49]. Above 10 K, all samples exhibit a similar increasing trend in heat capacity. However, below 10 K, the heat capacity curves diverge significantly. The highest values are observed for the low-density pure EG sample (#4-Chem$^{I}$), whereas the lowest values are recorded for the commercial Sinograf GFL05 sample, reflecting the impact of structural and density variations on thermal properties.

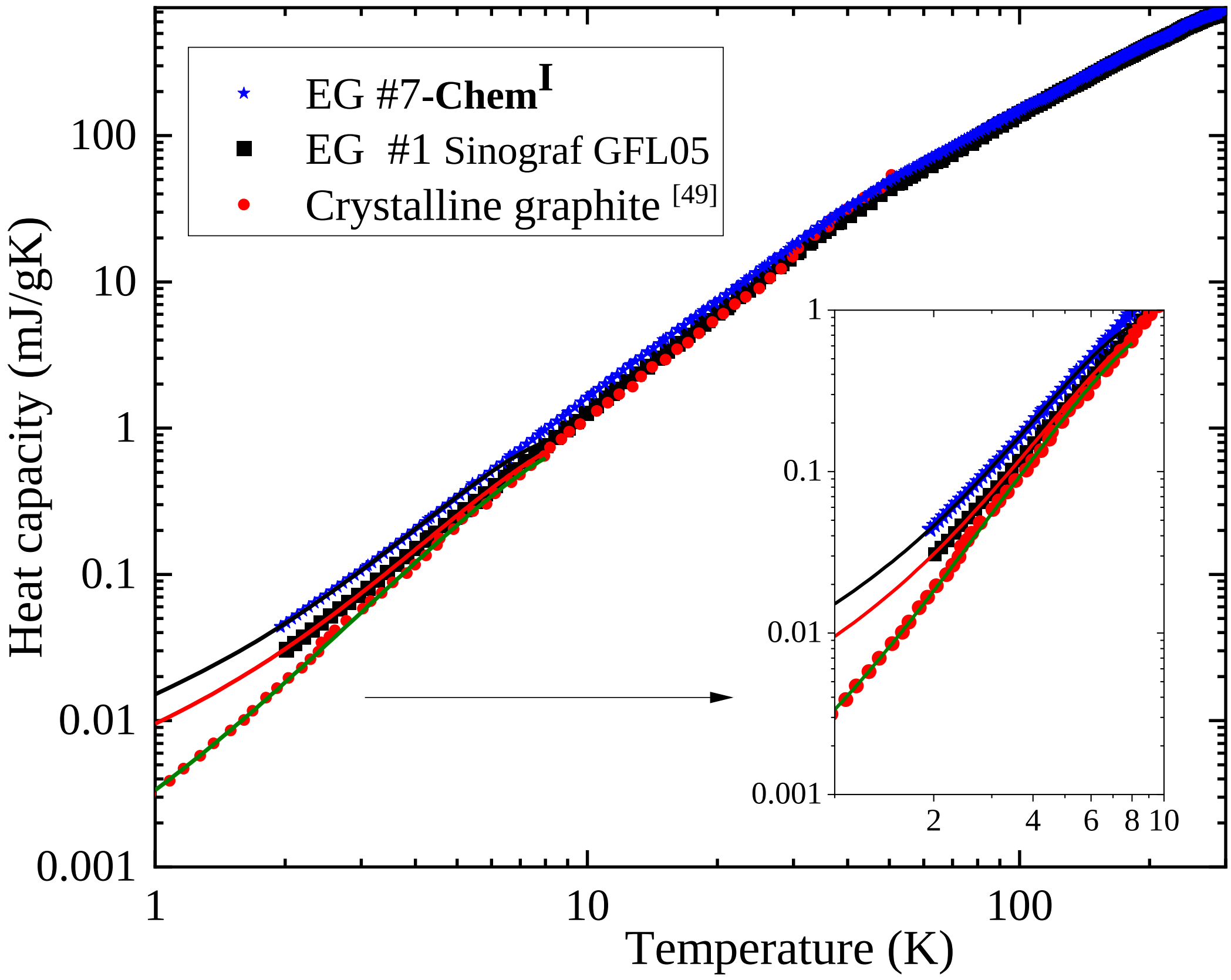

**Fig. 3** The temperature dependence of the heat capacity, different symbols indicate different pure carbon samples: ★ – EG #7-**Chem**[I]; ■ – EG Sinograf GFL05 and ● – crystalline graphite (highly oriented pyrolytic graphite, HOPG) [49]. The solid lines show the low-temperature approximation according to eq. (1) with fitting parameters in summarized in **Tab. 4**.

Temperature-dependent heat capacity measurements were subsequently performed on the elastically and plastically deformed samples, with the corresponding results summarized in Tables 4 and 5. Samples subjected to elastic–plastic deformation are denoted by the letter “p” at the end of their names.

The EG and EG–MWCNT composite samples represent flexible graphite materials with densities ranging from 0.23 g cm$^{-3}$ to 1.74 g cm$^{-3}$. Derived from expanded graphite, these materials retain a characteristic worm-like layered morphology (up to a density of ~ 0.4 g/cm$^3$) and a turbostratic microstructure arising from layer misorientation and structural defects. The average package thickness (coherent scattering domain size) $L_c$ (Table 3) is ~ 20 nm, with interlayer spacings of 3.35(4)–3.37(9) Å, reflecting a certain degree of structural disorder determined by specific synthesis conditions.

Overall, the heat capacity of both EG and Sinograf EG samples surpasses that of crystalline graphite, with the difference becoming increasingly pronounced at lower temperatures. Notably, this divergence appears to be independent of CNT concentration (see Fig. 4a).

Deformed samples exhibit a similar temperature dependence; however, their heat capacity is significantly lower in low-density composites, particularly in #10-**Chem^I^**-1%CNTs, #11-**EleChem^I^**-1%CNTs, and #4-**Chem^I^**, which demonstrate the most substantial deviations (see Fig. 4b).

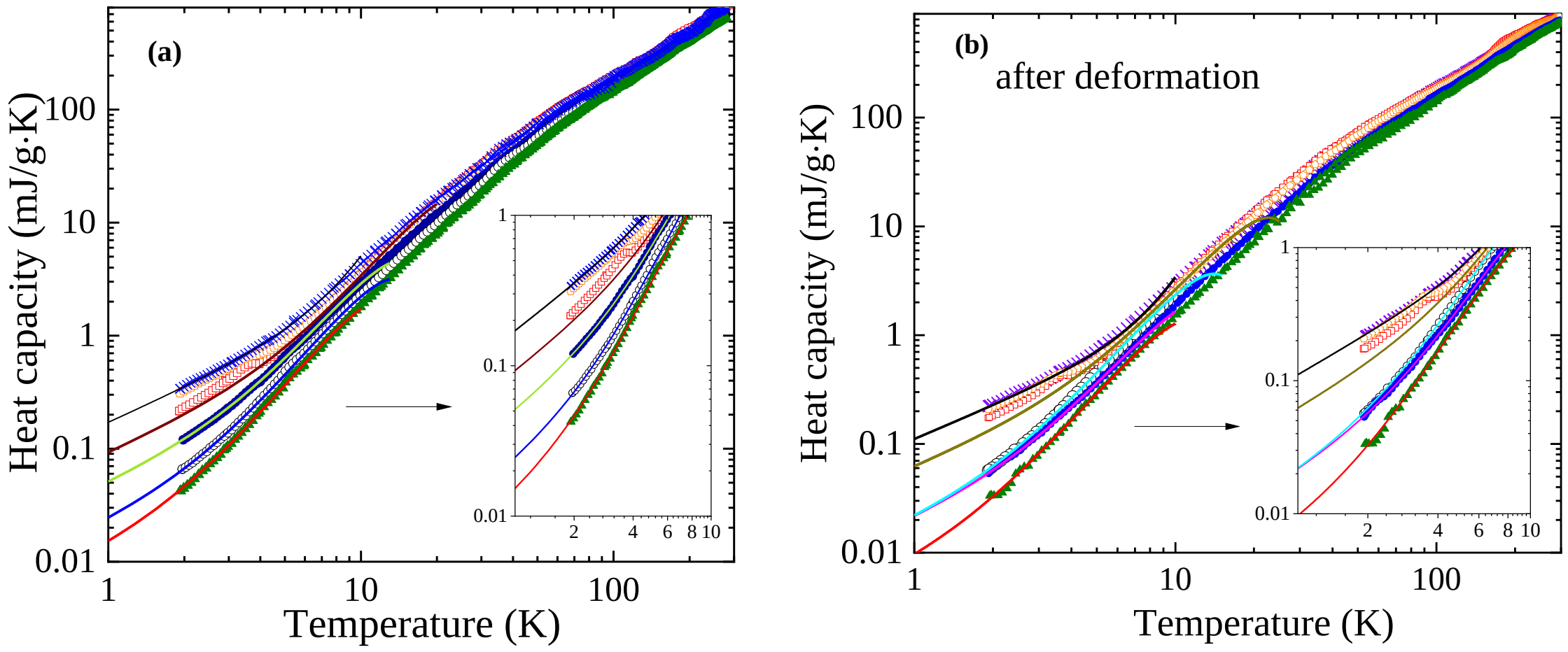

**Fig. 4** The temperature dependence of the heat capacity of EG-MWCNTs composites: a) pristine samples; b) after deformation. Symbols: × – #10-**Chem$^{I}$**-1%CNTs; ⬡ –#4-**Chem$^{I}$**; □ – #11-**EleChem$^{I}$**-1%CNTs; ● – #5-**Chem$^{II\text{-}III}$**; ○ – #9-**EleChem$^{I}$**-1%CNTs; ▲– #6-**EleChem$^{II\text{-}III}$**. The solid lines represent the fitting curves according to eq. (1), with fitting parameters $C_1$, $C_3$, $C_5$ in **Tab. 5**.

The representation of heat capacity data for EG samples, including deformed variants, as $C_p(T)/T$ versus $T^2$ (see Figs. 5a, 6a, and 6c) facilitates the application of a low-temperature polynomial approximation, given by [50]:

$$C_p(T) = C_1T + C_3T^3 + C_5T^5 \quad (1)$$

This three-term expansion is widely recognized for its applicability in describing the low-temperature behavior of both crystalline and amorphous bulk solids. Moreover, it exhibits strong agreement with experimental results for highly anisotropic and heterogeneous materials, particularly in carbon-based systems [30, 50].

**Table 4** Fitting parameters for the low-temperature approximation (Eq. 1) of heat capacity for EG samples, both pristine and deformed (pressed) variants, denoted by "p". Density ($\rho$) values are provided either by the material manufacturer or sourced from literature (see references in the text).

| Sample | $\rho$ | $C_1$ | $C_3$ | $C_5$ |
|---|---|---|---|---|
| | g/cm$^3$ | μJ/g·K$^2$ | μJ/g·K$^4$ | μJ/g·K$^6$ |
| **EG** | | | | |
| #4-**Chem$^{I}$** | 0.235 | 110 | 2.75 | -0.0038 |
| #4-**Chem**$^{I}$ p | | 60 | 2.47 | -0.0025 |
| #5-**Chem**$^{II-III}$ | 0.231 | 47.8 | 3.4 | -0.01 |
| #5-**Chem**$^{II-III}$ p | | 19.5 | 2.35 | -0.009 |
| #6-**EleChem**$^{II-III}$ | 0.357 | 12.6 | 2.7 | -0.011 |
| #6-**EleChem**$^{II-III}$p | | 7.4 | 2.3 | -0.011 |
| #7-**Chem**$^{I}$ | 0.73 | 12.4 | 2.78 | -0.024 |
| #7-**Chem**$^{I}$ p | | 13.5 | 2.4 | -0.013 |
| #8-**Chem**$^{I}$ | 1.3 | 12.4 | 2.8 | -0.024 |
| #8-**Chem**$^{I}$ p | | 13.5 | 2.4 | -0.013 |
| **EG Sinograf** | | | | |
| GFL05 | 0.7 | 7.5 | 2.05 | -0.013 |
| GFC025 | 1.0 | 9.7 | 2.3 | -0.011 |
| GFC10 | 1.0 | 11 | 1.97 | -0.010 |
| **HOPG** [49] | 2.0 | 1.37 | 2.0 | -0.0125 |

Equation (1) describes two fundamental contributions to heat capacity: a non-phonon term ($C_1T$) and a phonon component expressed through a temperature-series expansion, $C_{ph} = C_3T^3 + C_5T^5$. The linear term ($C_1T$) is typically attributed to either two-level systems (TLS) in disordered materials or an electronic contribution ($\gamma T$) in conductive systems.

For graphite and non-graphite carbon materials, the low-temperature heat capacity (below 2 K) can be effectively modeled using a simplified expression [7-18]:

$$C(T) = \gamma T + C_3T^3 \quad (2)$$

Here, $\gamma T$ accounts for electronic and defect-related contributions, while $C_3T^3$ represents the Debye phonon term, governing lattice vibrations [37, 51-53].

The heat capacity data for all EG and EG–MWCNTs composites exhibit strong agreement with the three-term approximation (Eq. 1) within the temperature range below 6 K. The corresponding fitting parameters (coefficients $C_1$, $C_3$, and $C_5$) are provided in Tables 4 and 5. The temperature dependence of heat capacity for various EG samples, plotted as $C_p/T$ versus $T^2$, is illustrated in Figs. 5a, 6a, and 6c, highlighting the consistency of the model with experimental results.

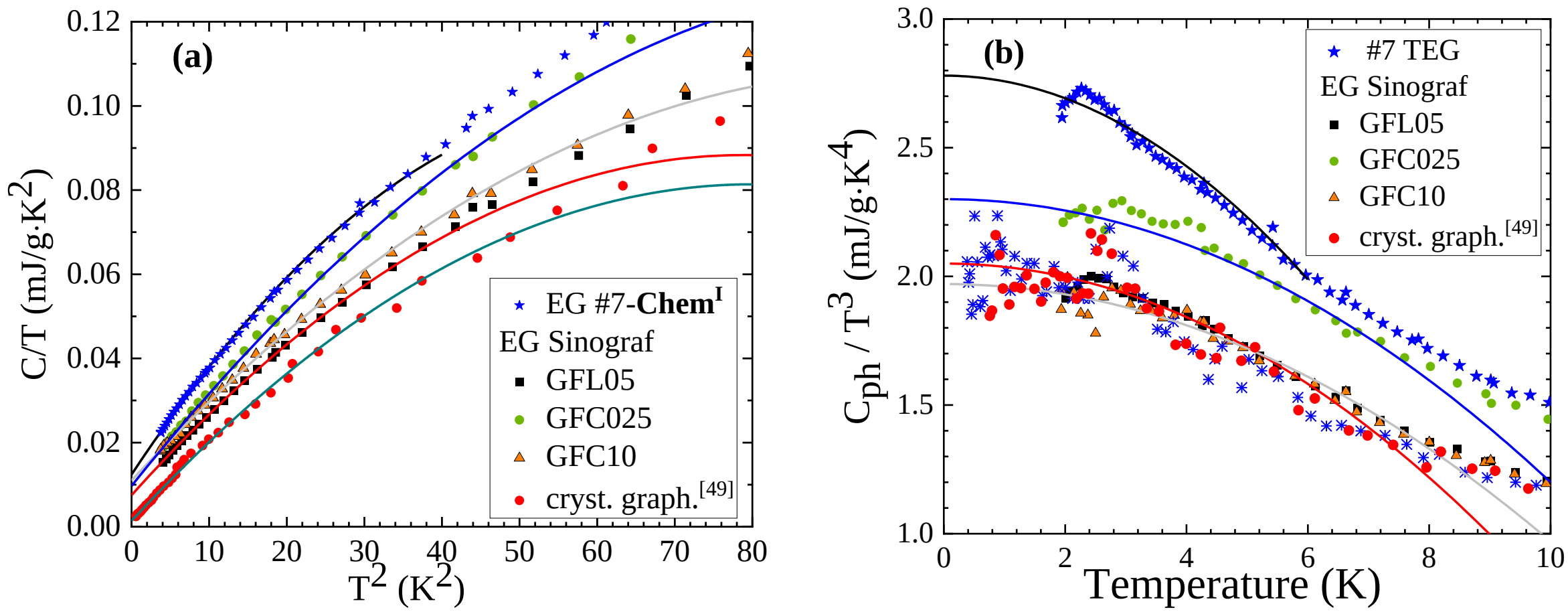

**Fig. 5** The temperature dependence of the heat capacity of different pure graphite samples in reduced coordinates: a) specific heat data plotted as $C_p/T$ vs $T^2$; b) Debye-reduced specific heat data $C_{ph}/T^3$ vs $T$. Solid lines show fitting curves according to eq. (1) with a fitting parameters from Tab.4.

**Table 5** Parameters for describing the heat capacity data for the EG-MWCNTs composites according to eq. (1).

| Sample | $\rho$ | $C_1$ | $C_3$ | $C_5$ |
|---|---|---|---|---|
| | g/cm³ | μJ/g·K² | μJ/g·K⁴ | μJ/g·K⁶ |
| **EG – 1% CNTs** | | | | |
| #9-**EleChem$^{I}$**-1%CNTs | 0.357 | 21.6 | 3 | -0.01 |
| #9-**EleChem$^{I}$**-1%CNTs p | | 19 | 3.1 | -0.0095 |
| #10-**Chem$^{I}$**-1%CNTs | 0.245 | 169 | 2.07 | 0.013 |
| #10-**Chem$^{I}$**-1%CNTs p | | 110 | 0.95 | 0.0135 |
| #11-**EleChem$^{I}$**-1%CNTs | 0.233 | 90 | 2.8 | -0.003 |
| #11-**EleChem$^{I}$**-1%CNTsp | | 52 | 2.47 | -0.0025 |
| #12-**Chem$^{I}$**-1%CNTs | 1.74 | 14.5 | 2.55 | -0.019 |
| #12-**Chem$^{I}$**-1%CNTs p | | 15.5 | 2.50 | -0.013 |
| **EG – 3% CNTs** | | | | |
| #13-**Chem$^{I}$**-3%CNTs | 0.966 | 11.8 | 2.66 | -0.022 |
| #13**Chem$^{I}$**-3%CNTs p | | 13.7 | 2.50 | -0.013 |
| #14-**Chem$^{I}$**-3%CNTs | 1.62 | 11.8 | 2.55 | -0.019 |
| #14-**Chem$^{I}$**-3%CNTs p | | 13.5 | 2.40 | -0.013 |

A notable observation is that both EG and EG–MWCNTs composites exhibit a linear coefficient ($C_1$) that exceeds that of crystalline graphite by more than an order or even two orders of magnitude. This suggests that the dominant contribution to the linear term is non-electronic in origin.

Elastic–plastic deformation in low-density samples ($\rho$ < 0.3 g cm$^{-3}$) results in a pronounced reduction of the linear heat capacity term $C_1T$, as illustrated in Fig. 7. Compression at 2 MPa induces reversible (elastic) and irreversible (plastic) changes within the worm-like layered structure of porous samples ($\rho \approx 0.2–0.4$ g/cm$^3$) due to mechanical stress acting on weak elements of the structural 'framework' [35, 47].

These irreversible (plastic) transformations have a strong impact on the low-temperature heat capacity of expanded graphite. EG, a highly porous and low-density form of graphite, initially exhibits elastic deformation under small stresses, recovering its shape upon unloading. However, when the applied stress exceeds a critical threshold, plastic deformation occurs, leading to permanent structural rearrangements.

The plastic deformation effect is not observed in higher-density samples because their compaction into foils already involves much higher pressures than 2 MPa [35, 36, 47].

The coefficient $C_3$ represents the Debye term, whereas coefficient $C_5$ reflects the influence of non-linear phonon dispersion [30]. In elastically isotropic crystals, coefficient $C_5$ is typically positive [54], resulting in the well-known “hump” in heat capacity when plotted as $C_{ph}/T^3$ [55]. However, in carbon-based and strongly anisotropic materials [30], this trend may be reversed.

The findings for EG and EG–MWCNTs samples strongly support the interpretation of a negative coefficient $C_5$ as an indicator of non-linear (flexural) phonon dispersion. Among the samples characterized by pronounced anisotropic layered structures, the $C_{ph}/T^3$ hump is completely absent, and $C_5$ is negative (see Figs. 5b, 6b, and 6d). Notably, the characteristic $C_{ph}/T^3$ hump with a maximum near 9 K is observed only in two samples—#10-**Chem$^{I}$**-1%CNTs and #10-**Chem$^{I}$**-1%CNTs-p—which also display high oxygen and sulfur impurity levels, likely intercalated within the EG layers (p.102 [35]).

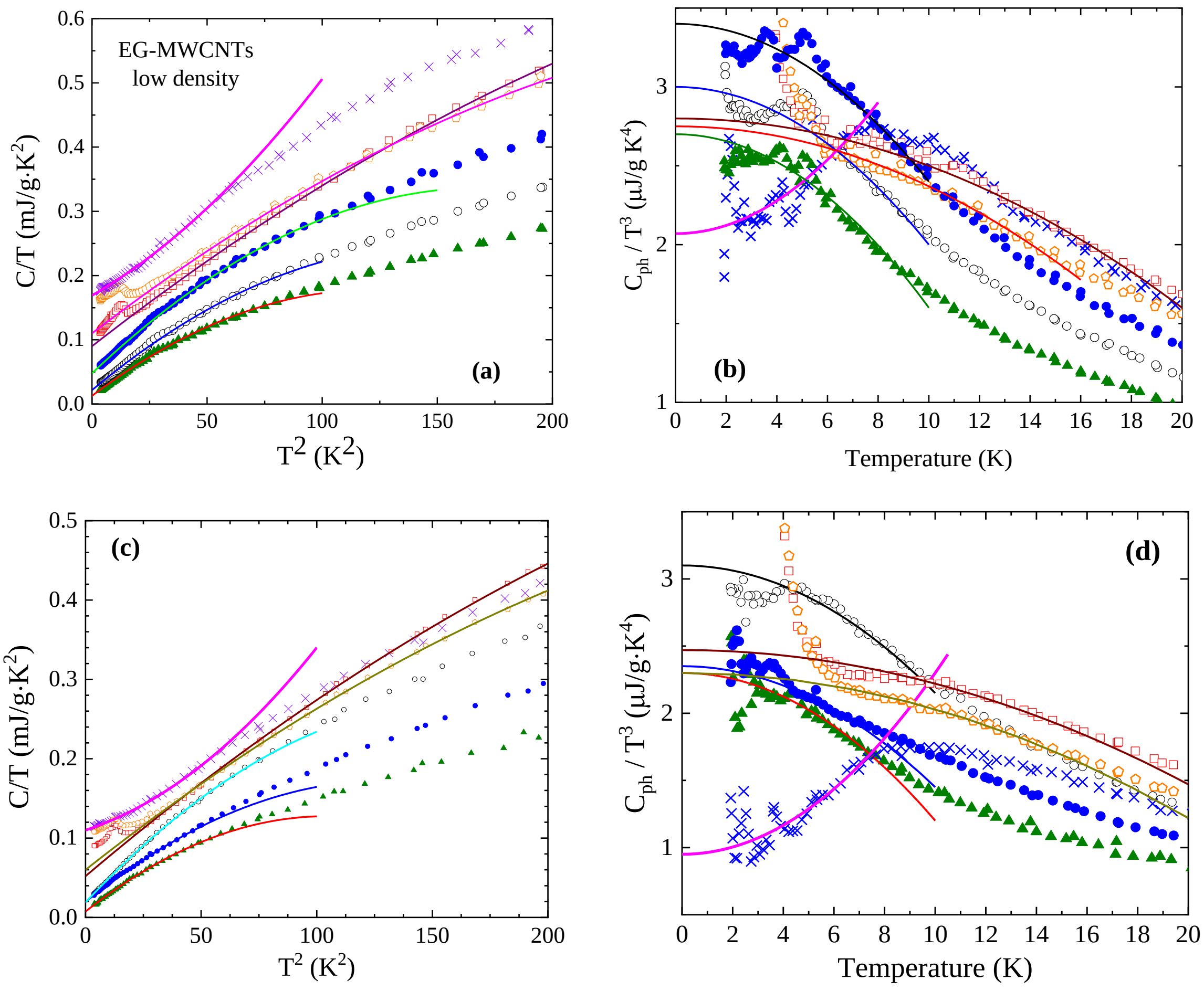

**Fig. 6** Heat capacity of EG-MWCNTs composites in reduced coordinates: (a) and (b) pristine samples; (c) and (d) pressed samples. Symbols: × – #10-**Chem**$^{\mathbf{I}}$-1%CNTs; ⬡ –#4-**Chem**$^{\mathbf{I}}$; □ – #11-**EleChem**$^{\mathbf{I}}$-1%CNTs; ● – #5-**Chem**$^{\mathbf{II\text{-}III}}$; ○ – #9-**EleChem**$^{\mathbf{I}}$-1%CNTs; **▲– #6-EleChem**$^{\mathbf{II\text{-}III}}$. The solid lines represent the fitting curves according to eq.(1), the corresponding fitting parameters $C_1$, $C_3$, $C_5$ are presented in Table 5.

The Debye coefficient ($C_3$) in EG and EG–MWCNTs composites is slightly higher than that of crystalline graphite, reflecting differences in lattice dynamics due to the increased disorder and heterogeneous microstructure of the composites.

A particularly significant finding is the negative value of $C_5$ observed in polycrystalline layered graphite structure of EG composites. This behavior corresponds to the quadratic dispersion relation ($\omega \sim k^2$) in the long-wavelength limit, which deviates from the linear dispersion ($\omega = ks$) typically associated with acoustic phonon modes at higher quasi-momentum values [30].

In free-standing single-layer and bilayer graphene, theoretical studies indicate that at low frequencies, out-of-plane flexural (ZA) phonons follow a quadratic dispersion relation ($\omega \propto k^2$), resulting in a linear temperature dependence of specific heat ($C \propto T$). Conversely, the in-plane transverse (TA) and longitudinal (LA) acoustic phonons exhibit linear dispersion ($\omega \propto k$), leading to a quadratic temperature dependence of specific heat ($C \propto T^2$) [56-58].

For quasi-two-dimensional anisotropic systems, such as exfoliated graphite, atomic displacements perpendicular to basal planes (out-of-plane motions) are significantly larger than in-plane displacements [59-61]. This results in the dominance of the flexural acoustic (ZA) mode, whose anomalous dispersion manifests as a negative coefficient $C_5$ in the heat capacity analysis.

The temperature dependence of $C_{ph}/T^3$ across various EG samples is similar to that observed in graphite—remaining nearly constant below 2 K before gradually decreasing at higher temperatures. This trend is a hallmark of disordered layered carbon systems, primarily driven by the anisotropic vibrational dynamics within their crystal structures, particularly the pronounced disparity between interlayer and intralayer bonding forces.

Consistent with previous findings for disordered single-walled and multi-walled carbon nanotubes [30], the characteristic $C_{ph}/T^3$ hump is absent in most EG samples. This lack of a hump reinforces the interpretation that a negative coefficient $C_5$ signifies non-linear phonon dispersion, with out-of-plane acoustic flexural modes dominating vibrational behavior.

The effect of elastic–plastic deformation on the heat capacity is emphasized in Fig. 7, which compares the coefficient $C_1$ values for pristine and deformed samples.

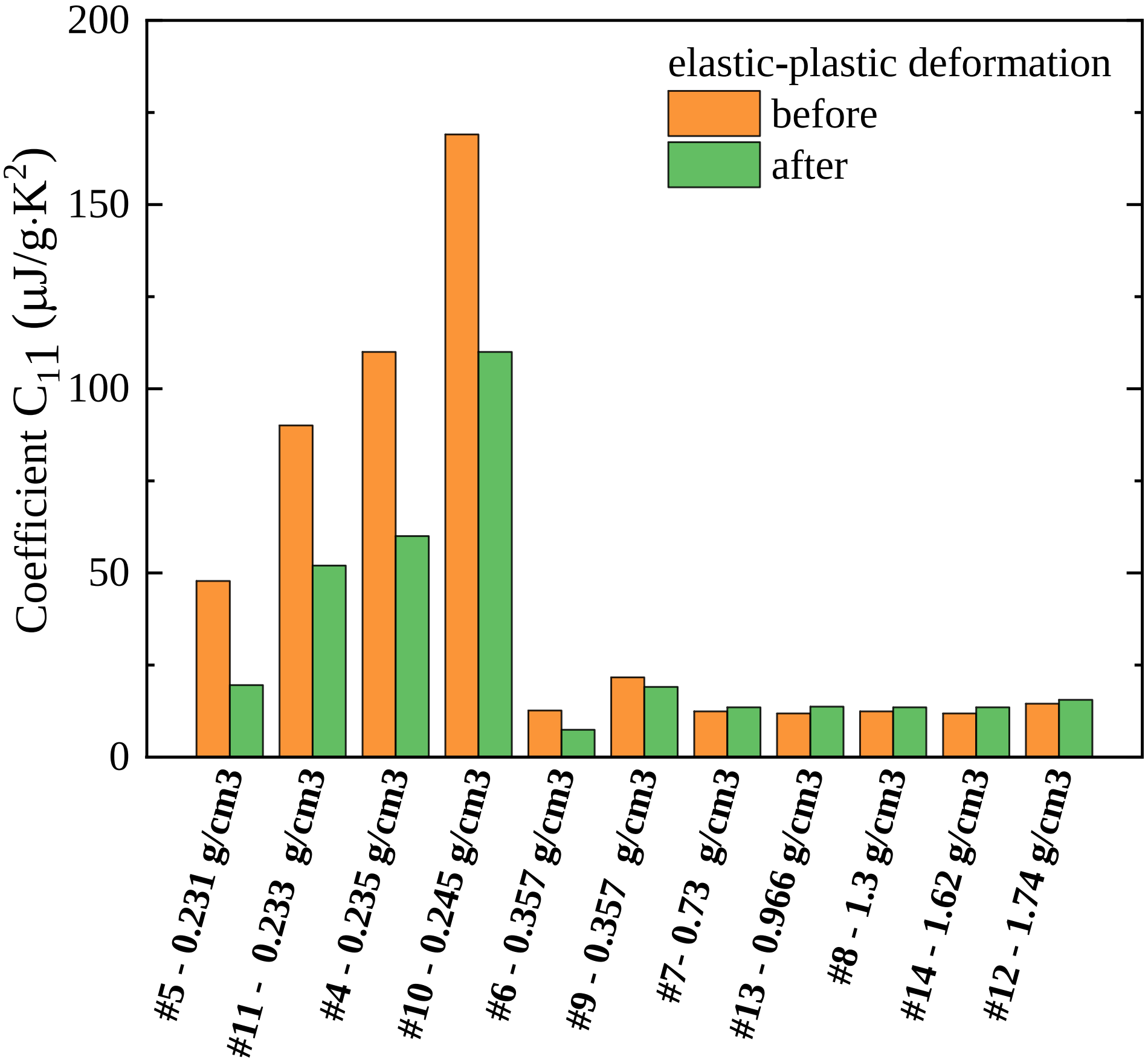

**Fig. 7** Effect of elastic–plastic deformation on the low-temperature heat capacity. Comparison of the linear term (coefficient $C_1$) for undeformed and deformed samples.

For low-density samples of pure EG and EG composites containing 1% CNTs, elastic-plastic deformation leads to a significant reduction in $C_1$, decreasing by nearly 30%. In contrast, high-density samples of both pure EG and EG–MWCNTs composites exhibit a moderate increase in coefficient $C_1$, remaining below 10%. This contrasting behavior suggests that the impact of elastic-plastic deformation on heat capacity is closely linked to the material's initial density and microstructure. A clear correlation has been identified between the linear heat capacity contribution (coefficient $C_1$) and structural characteristics derived from Raman spectroscopy, XRD, and EDS analyses, as illustrated in Fig. 8:

- Raman spectroscopy data (Table 2) indicate that the $I_{2D}/I_G$ ratio, which quantifies defect levels in multilayer graphite, aligns with the highest values of the linear heat capacity term (coefficient $C_1$).
- The inclusion of multi-walled carbon nanotubes (MWCNTs) up to 3 wt.% does not significantly affect the heat capacity, which emphasizes that structural disorder induced by hydrolysis of intercalated compounds and heat treatment under non-equilibrium rapid heating conditions in the thermal shock regime, rather than the nanotube concentration, are the main determinants of coefficient $C_1$.

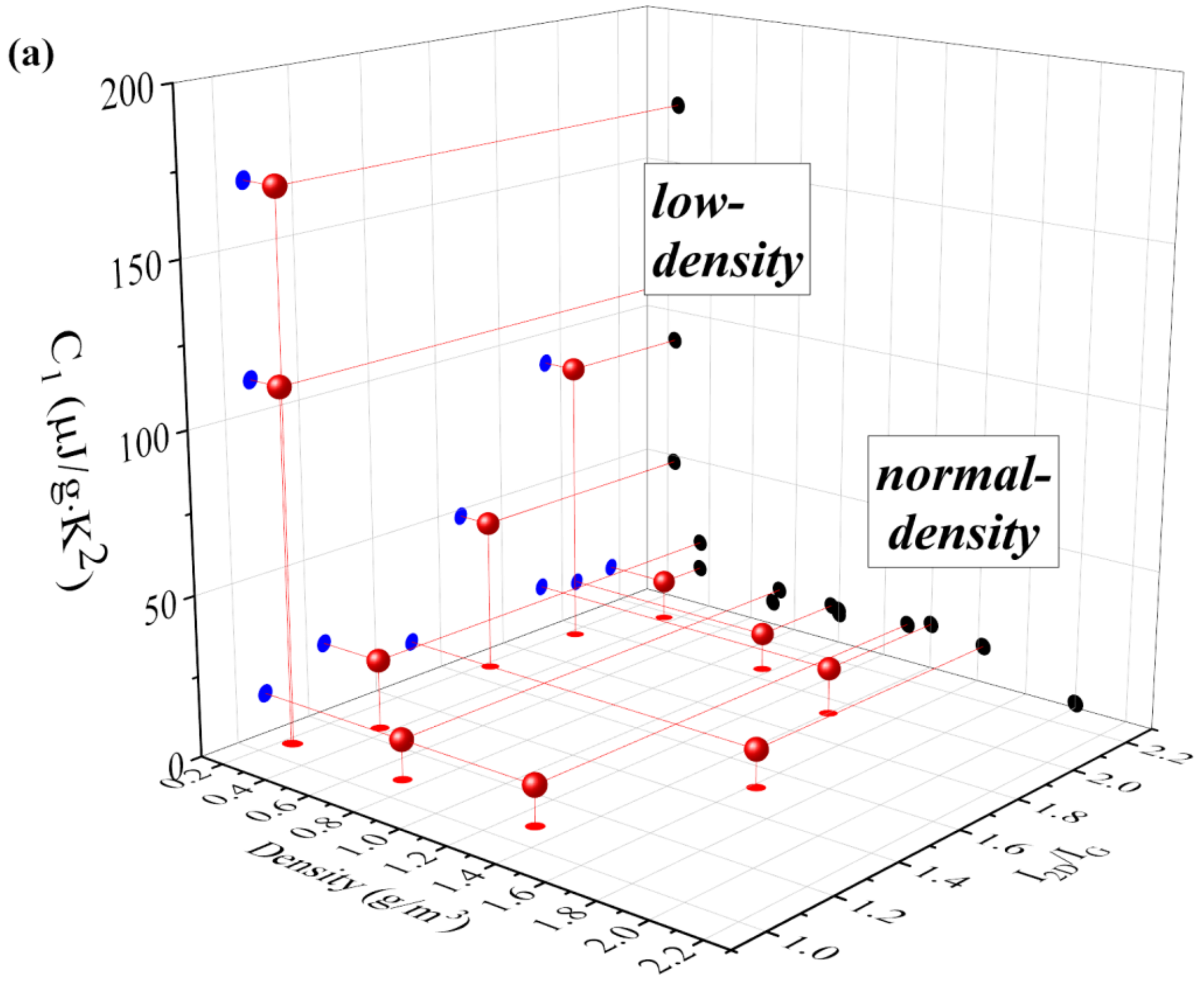

**Fig. 8** Dependence of the $C_1$ fitting parameter: a) $C_1$ vs density vs. $I_{2G}/I_G$.

These findings highlight the low-temperature heat capacity of multilayer carbon materials as a sensitive probe of their defect structure and inhomogeneity. The structural imperfections in expanded graphite (EG) and EG–MWCNTs composites contribute significantly to two distinct components of heat capacity: the anomalous linear term and the phonon contributions.

Loosely packed, defect-rich graphite structures, typically associated with low density, exhibit higher heat capacity compared to their more compact or crystalline counterparts. The phonon heat capacity, defined as $C_{ph} = C(T) - C_1T$, with the linear term $C_1T$ subtracted, is presented in Fig. 5b and 7b as $C_{ph}/T^3$ versus temperature. This characteristic behavior in disordered EG systems is attributed to their unique vibrational dynamics, shaped by the pronounced anisotropy in interatomic interactions within their layered crystal structure.

Similar to disordered SWCNTs and MWCNTs [30], no $C_{ph}/T^3$ hump is observed across most EG samples, reinforcing the presence of nonlinear phonon dispersion and further supporting the negative $C_5$ parameter as a signature of flexural acoustic phonon dominance.

## 4. Discussion

The research demonstrates that the low-temperature heat capacity of expanded graphite (EG) is primarily governed by low-frequency out-of-plane phonons exhibiting quadratic dispersion relations. These phonon modes, characteristic of two-dimensional layered systems, dominate vibrational behavior at cryogenic temperatures. Compared to bulk crystalline graphite, EG consistently shows higher heat capacity at low temperatures, primarily due to its increased defect density, structural disorder, and reduced coherence length between layers.

Experimental findings across multiple studies confirm that the low-temperature heat capacity of graphite-based materials is highly sensitive to crystallite size, stacking order, chemical composition, and density. In samples, the three-term fitting model (Eq. 1) accurately describes experimental heat capacity data up to 6 K, with each term correlating to distinct physical mechanisms. Specifically, the linear coefficient $C_1$ is predominantly attributed to structural defects and disordered regions [7-18, 51].

This behavior is consistently observed in both pure EG and EG–MWCNTs composites, with the addition of multi-walled carbon nanotubes (up to 3 wt%) exerting only a minor influence. Furthermore, the $I_{2D}/I_G$ ratio obtained from Raman spectroscopy, an indicator of defect density in multilayered graphite structures, follows a similar trend with coefficient $C_1$, reinforcing the link between vibrational heat capacity and structural imperfections.

The connection between disordered networks and thermal properties was first conceptualized by Franklin [62], who proposed that graphite-like structures consist of platelike crystalline domains interconnected by disordered carbon bridges. In our case, elastic-plastic deformation likely disrupts some of these bridges, leading to irreversible (plastic) changes within the worm-like layered structure of porous samples ($\rho \approx 0.2–0.4$ g/cm$^3$) end a reduction in the linear $C_1$ coefficient.

Compression at 2 MPa induces reversible (elastic) and irreversible (plastic) changes within the worm-like layered structure of porous samples ($\rho \approx 0.2–0.4$ g/cm$^3$) due to mechanical stress acting on weak elements of the structural 'framework' [35, 47].

Notably, we consistently observe a negative $C_5$ coefficient across most expanded graphite (EG) samples, indicating quadratic dispersion ($\omega \sim k^2$) for out-of-plane flexural (ZA) modes, characteristic of two-dimensional layered systems. This deviates from the linear $\omega = sk$ dispersion typical of in-plane acoustic modes, reflecting the anisotropic structure of EG, where out-of-plane atomic vibrations dominate at long wavelengths.

The low-frequency phonon dispersion in EG can be effectively described by the equation [58, 61]:

$$\omega^2(k) = (s\,k)^2 + f k^4, \qquad (3)$$

where $s$ represents the in-plane sound velocity, and $f$ is a softening parameter related to the flexural rigidity of the graphene layers.

Wirtz and Rubio [63] revisited the phonon dispersion of graphite using first-principles calculations, demonstrating excellent agreement with experimental observations. One major consequence of Eq. (3) is the absence of excess heat capacity in the temperature range where non-layered crystals and glasses typically exhibit a low-temperature hump (or "boson peak") in their heat capacity curves.

The absence of this characteristic hump has also been reported in other low-dimensional or disordered carbon materials, including non-crystalline carbons [13, 14] nanotubes [30], certain molecular crystals [64 -70] and biomorphic SiC/Si composites [71]. However, the detailed dynamics of flexural phonons in these materials remain an open and actively researched topic.

**Conclusions**

Structural characterization via XRD, Raman spectroscopy, and EDS revealed significant variations in layer stacking, defect concentration, and elemental impurities, particularly in chemically oxidized samples. Heat capacity measurements demonstrated that structural disorder and MWCNTs integration led to an increase in specific heat.

The temperature dependence of heat capacity was analyzed using a three-term power-series expansion ($C_1T$, $C_3T^3$, $C_5T^5$). Low-density and highly disordered samples exhibited a negative $C_5$ coefficient, suggesting enhanced flexural phonon dispersion. The linear term ($C_1T$) was significantly higher in EG and EG–MWCNTs composites than in crystalline graphite, correlating strongly the density of material.

A negative $C_5T^5$ coefficient across most samples confirms the dominance of quadratic flexural phonon modes, characteristic of anisotropic layered structures like EG. However, in samples containing intercalated impurities (e.g. oxygen, sulfur), a residual phonon hump remains visible.

Elastic-plastic deformation reduced the linear heat capacity term in low-density composites, indicating increased structural compaction. Across all densities, EG and EG–MWCNTs composites consistently exhibited higher heat capacity than crystalline graphite, with chemically oxidized, low-density samples showing the highest values. The typical $C/T^3$ "hump" feature was suppressed in disordered, low-density composites, suggesting MWCNTs disrupt phonon confinement and scattering mechanisms.

These findings emphasize the importance of density and disorder in tailoring EG-MWCNTs thermal properties, providing valuable insights for thermal management, energy storage, and electronic applications.

**Conflict of interest**

The authors declare no competing interests.

**Data availability statement**

Data will be made available from the corresponding author upon reasonable request.

**Authors contribution**

A.I. Krivchikov conceived the study, developed the theoretical framework, performed the computations, supervised the research, and wrote the manuscript.
A. Jeżowski co-conceived the work, supervised the research, and contributed to project coordination.
M.S. Barabashko supervised data analysis, verified analytical methods, and contributed to manuscript preparation.
G. Dovbeshko co-conceived the study, performed Raman experiments, analyzed the data, and contributed to writing.
D.E. Hurova performed XRD measurements and validated the analytical procedures, supported by N.N. Galtsov.
V. Boiko conducted Raman measurements.
Yu. Sementsov fabricated the samples, verified EDS analysis, and contributed to manuscript writing.
A. Glamazda verified Raman analytical methods.
V. Sagan supervised the project and validated analytical methods.
Yu. Horbatenko and O. Korolyuk verified analytical methods and contributed to manuscript preparation.
O. Romantsova and D. Szewczyk performed heat capacity measurements; D. Szewczyk also assisted in data analysis and project supervision.

**Funding**
This work was partly supported by the National Research Foundation of Ukraine (Grant 2023.03/0012) and National Science Centre Poland (Grant 2022/45/B/ST3/02326 and 2025/09/X/ST3/00153). One of us (O.A.K.) expresses deep gratitude to the International Visegrad Fund for partial financial support of the research within Project No 62410077.